\documentclass[%
 reprint, 
 amsmath,amssymb,
 aps,
floatfix
]{revtex4-1}

\usepackage{graphicx}
\usepackage{dcolumn}
\usepackage{bm}
\usepackage{hyperref}
\usepackage{upgreek}
\usepackage{xcolor}

\begin{document}

\title{Electroneutrality Breakdown in Nanopore Arrays}
\author{J. Pedro de Souza$^1$}%

\author{Amir Levy$^2$}%

\author{Martin Z. Bazant$^{1,3}$}%
\affiliation{%
 $^1$Department of Chemical Engineering, Massachusetts Institute of Technology, Cambridge, Massachusetts 02139 USA
}%
\affiliation{%
 $^2$Department of Physics, Massachusetts Institute of Technology, Cambridge, Massachusetts 02139 USA
}%
\affiliation{%
 $^3$Department of Mathematics, Massachusetts Institute of Technology, Cambridge, Massachusetts 02139 USA
}%

\begin{abstract}
    The electrostatic screening of charge in one-dimensional confinement leads to long-range breakdown in electroneutrality within a nanopore. Through a series of continuum simulations, we demonstrate the principles of electroneutrality breakdown for electrolytes in one dimensional confinement. We show how interacting pores in a membrane can counteract the phenomenon of electroneutrality breakdown, eventually returning to electroneutrality. Emphasis is placed on applying simplifying formulas to reduce the multidimensional partial differential equations into a single ordinary differential equation for the electrostatic potential. Dielectric mismatch between the electrolyte and membrane, pore aspect ratio, and confinement dimensionality are studied independently, outlining the relevance of electroneutrality breakdown in nanoporous membranes for selective ion transport and separations.
\end{abstract}
\maketitle
\section{Introduction}
The conduction of ions in nanochannels forms the basis of electrical signalling in biology ~\cite{hille2001ion, eisenberg1996computing, doyle1998structure, catterall2000ionic, sakmann2013single, Maffeo2012} and of promising technologies in desalination ~\cite{shannon2010science, goh2016nano}, ionic separations \cite{lu2020efficient, li2020strong}, and nanofluidic transistors ~\cite{karnik2007rectification, karnik2005electrostatic, daiguji2010ion}. As ions become confined to nanopores, they experience strong chemical and electrostatic interactions with the pore walls, leading to membrane selectivity based on charge or chemical interactions. Nanoporous membranes can even enter the regime where the double layers emanating from each charged surface begin to overlap, leading to strong electrokinetic coupling of fluid flow, electric field, and ionic fluxes ~\cite{gross1968membrane, peters2016analysis, biesheuvel2016analysis, alizadeh2017multiscale, qiao2005scaling, qiao2005surface}. Understanding the electrostatics of charges in confinement is crucial to determining the flux, selectivity, and driving force relationship for engineering applications ~\cite{elimelech2011future,faucher2019critical} and for understanding biological pore systems ~\cite{Branton2008,HAQUE201356, gillespie2002coupling, chen1997permeation}.

The unique physics of ionic screening in one dimensional confinement leads to the phenomenon of electroneutrality breakdown, where the number of counter-charges within a pore does not perfectly counterbalance the number of fixed charges on the pore walls ~\cite{levy2020breakdown}. One dimensional (1D) confinement refers to confinement onto a line, for example, in a cylindrical nanopore connecting two reservoirs of fixed concentration. In essence, electroneutrality breakdown signifies that a fraction of the electric field must escape through the pore walls into the dielectric matrix constituting the membrane. The screening charge does not exist locally within the pore, but rather is distributed over the membrane surface in the reservoirs, outside of the membrane domain. Uniquely, in 1D confinement, the loss of electroneutrality can extend to macroscopic scales (beyond length $L=10$ $\mu$m), since for strong confinement $\kappa_D R\rightarrow 0$ the system only approaches electroneutrality as $\log(L/R)\rightarrow \infty$, where $L$ is the length of a pore, $R$ is its radius, and $\kappa_D$ is the inverse Debye length. The long-range breakdown of local electroneutrality in 1D confinement is surprising, and many models of electrokinetic phenomena have assumed pore-wide electroneutrality, even in the limit of strong double layer overlap ~\cite{peters2016analysis, secchi2016scaling, biesheuvel2016analysis}.

Here, we perform continuum simulations of 1D nanopores using COMSOL Multiphysics, to confirm the occurrence of electroneutrality breakdown. We show that the results for the screening charge within an isolated 1D pore can be captured quantitatively with analytical formulas. Furthermore, we perform simulations of a periodic array of channels, a multipore membrane, with varying spacing between the channels. There, we find how the interactions of closely-spaced channels can lead to the return of electroneutrality in the system. In the limit of strongly interacting channels, the variations in potential in the axial direction dominate the distribution of ionic charges within the membrane. Effectively, when channels are too close to each other, the electric field lines cannot emanate through the membrane domain.

The continuum simulation results show the ensemble interactions of the channels with each other play a role in ionic conduction through nanochannels. In membrane applications, the interactions mean that the greatest ionic selectivity and per-channel-conductivity can be achieved when channels are close together for the regime of strong double layer overlap.  Further, the results point to ensemble interactions between 1D-confined channels, and the importance of electric field spilling into the dielectric matrix when channels are isolated. The competition between channel interactions and electroneutrality breakdown ultimately affects the conductance and selectivity behavior of arrays of nanochannels in the low concentration limit.

\begin{figure*}[!htbp] \label{fig:fig1}
\includegraphics[width=1\linewidth]{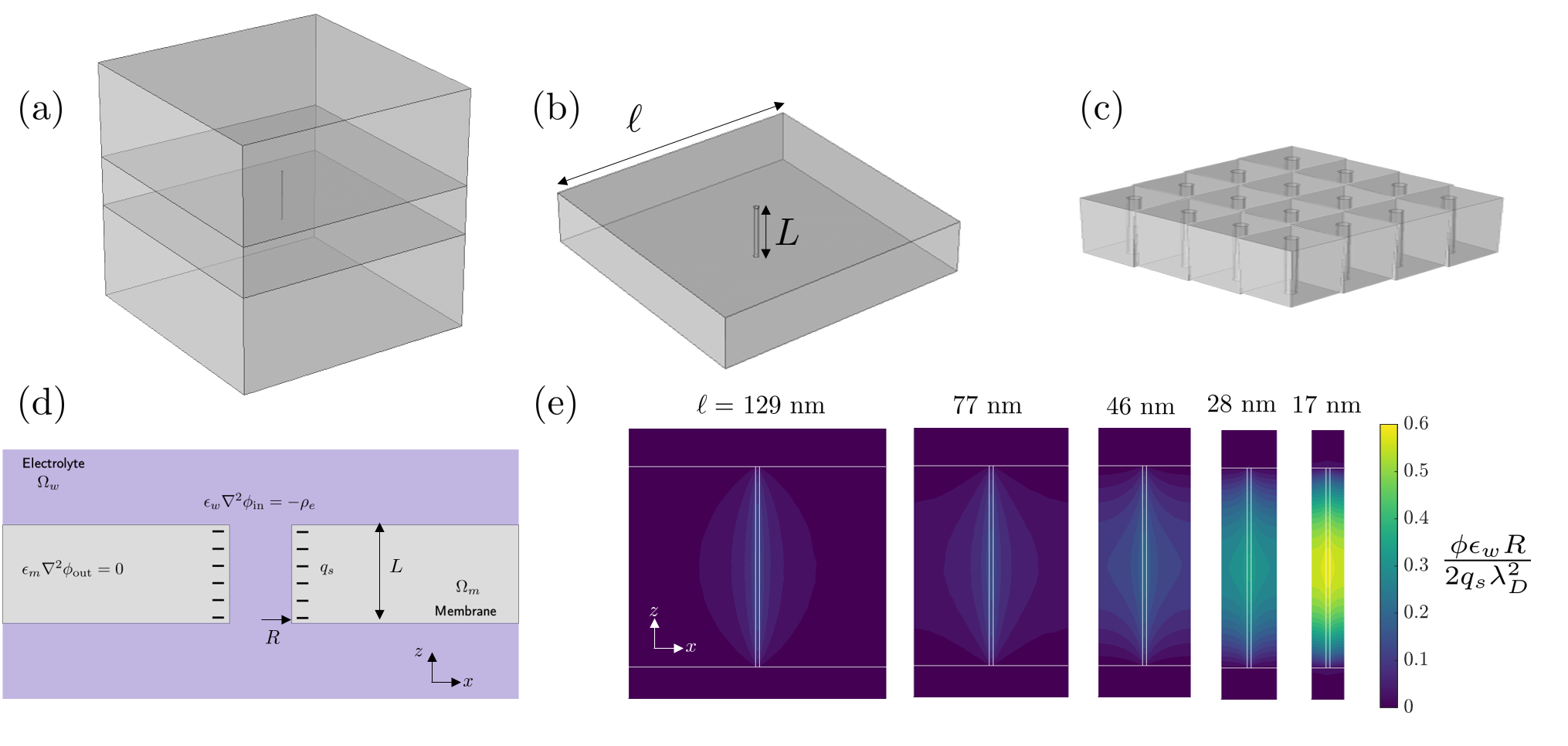}
\caption{The simulation configuration in COMSOL (a) with the membrane, pore, and electrolyte (b) for an isolated pore that does not feel its periodic neighbors and (c) a periodic arrangement of pores on a square lattice. (d) A cross-section of the system through the center of the cylindrical pore is shown to describe the domains and equations applied in the simulations. (e) A sample of continuum simulation results showing the progress from non-interacting to interacting channels for small charge densities. For all channels, the channel radius is 1 nm and the length is 100 nm. The membrane dielectric constant is $\epsilon_m=10\epsilon_0$ and the electrolyte dielectric constant is $\epsilon_w=80\epsilon_0$. The salt concentration is 1 mM. In order from left to right, the spacing between the channel centers $\ell$ is 129 nm, 77 nm, 46 nm, 28 nm, and 17 nm. Also in order, the amount of charge within the pore versus on the channel walls, $\mid Q_\mathrm{in}/Q_\mathrm{out}\mid$ is: 12\%,13\%, 17\%, 26\%, and 43\%. As the channels become closer together and more interacting, the system returns closer to electroneutrality, which is evidenced by the higher $\phi$ values within the pore. } 
\end{figure*}

\section{Theoretical framework}
\subsection{Outline of equations}
As explained in more detail in Appendix A, the ionic flux and selectivity out of equilibrium can be assumed to be related to the solution of the Poisson-Boltzmann (PB) equation in equilibrium, assuming that both reservoirs on each side of the membrane have the same concentration of electrolyte and fluid flow is neglected ~\cite{gross1968membrane, peters2016analysis}. As emphasized in Fig. 1, the simulation is composed of a square membrane domain, $\Omega_m$, with side length $\ell$, through which a cylindrical pore of radius $R$ and length $L$ connects two reservoirs of fixed concentration. The reservoirs and pore constitute the electrolyte domain, signified as $\Omega_w$, where $w$ is chosen to signify water.  

As exhibited in Fig. 1(a-d), the system of equations being solved in the electrolyte domain $\Omega_w$ is:
\begin{equation}
    \epsilon_w\nabla^2\phi=-\rho_e
\end{equation}
with dielectric constant $\epsilon_w$, electrostatic potential $\phi$, and charge density $\rho_e$ where the ionic concentrations are Boltzmann-distributed:
\begin{equation}
    \rho_e=\sum_i z_i e c_{i,b}\exp\left(-\frac{z_ie\phi }{k_B T}\right),
\end{equation}
where $z_i$ are the ion valencies, $e$ is the elementary charge, $c_{i,b}$ is the bulk reservoir concentration for ion $i$, $k_B$ is the Boltzmann constant, and $T$ is the absolute temperature. Here we neglect any packing ~\cite{bazant2009towards}, correlation ~\cite{Bazant2011}, or charge regulation ~\cite{trefalt2016charge} effects in our model.
We assume a 1:1 solution of salt with concentration $c_0$ in units of number density, such that the PB equation is reduced to: 
\begin{equation}
    \epsilon_w\nabla^2\phi=2 e c_0 \sinh\left(\frac{e\phi }{k_B T}\right)
\end{equation}

In the membrane domain, we solve the Laplace equation:
\begin{equation}
    \epsilon_m\nabla^2\phi=0
\end{equation}
assuming that the membrane is a perfect dielectric material with dielectric constant $\epsilon_m$. 

At the membrane/electrolyte domain interface, Maxwell's equation is enforced:
\begin{equation}
\mathbf{\hat{n}}\cdot\left[-\epsilon_m\nabla\phi+\epsilon_w\nabla \phi\right]\Bigg\rvert_s=q_s 
\end{equation}
where $\mathbf{\hat{n}}$ is the unit normal pointing from the electrolyte to the membrane domain and $q_s$ is the surface charge density. To isolate the electrostatic potential variations due to fixed charge on the channel walls, $q_s$ is assumed to be zero on the membrane/electrolyte reservoir interfaces, but is nonzero at the pore walls. 

At the lateral boundaries of each cell, symmetry conditions are applied; namely, the electric field at the boundary is zero $\mathrm{\hat{n}}\cdot \nabla \phi=0$. The boundary conditions are identical to assuming a periodic array of channels with regular spacing, consisting of unit cells identical to the simulation box, as illustrated in Fig. 1(c). At the top and bottom boundaries of the simulation box, Dirichlet conditions are applied $\phi=0$. A useful check to ensure that the simulation box is large enough is to ensure that all the integrated charges in the electrolyte domain are equal and opposite to the integrated amount of fixed charges on the pore walls. For this study, a reservoir height of 20 $\lambda_D$ is sufficient to meet this criterion, where the Debye length, $\lambda_D$ is given by:
\begin{equation}
    \lambda_D={\kappa_D}^{-1}=\sqrt{\frac{\epsilon_w k_B T}{2 e^2 c_0}}
\end{equation}

\begin{figure*}[!htbp] \label{fig:fig2}
\includegraphics[width=1\linewidth]{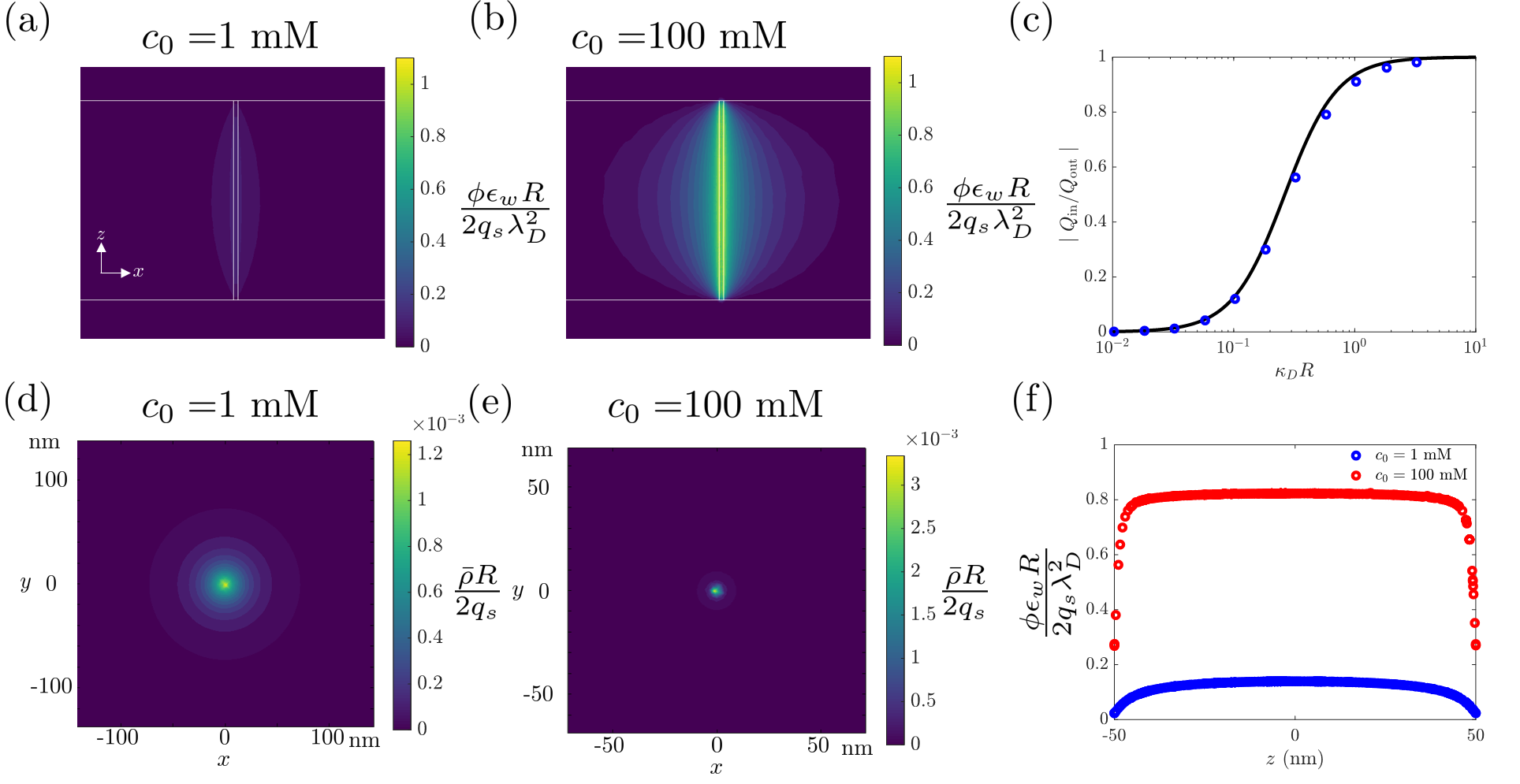}
\caption{The results for an isolated pore with overlapping and non-overlapping double layers within the pore.  (a) plot of the potential profile for overlapping double layers at $c_0=1 $ mM ($\kappa_DR\approx0.1$) (a) versus less overlapping double layers at $c_0=100$ mM ($\kappa_DR\approx1$)(b). The same parameters are used as in Fig. 1(e), except the center to center distance between pores is 500 nm such that the periodic channels are not interacting. For (a) 12\% of the charge is contained within the pore, whereas for (b) 91\% of the charge is contained within the pore.(c) The charge within the pore versus on the pore walls as a function of $\kappa_D R$ for the same channel in parts (a) and (b). Electroneutrality breakdown occurs in the region of strong double layer overlap $\kappa_D R\rightarrow 0$. For (c) the markers are the COMSOL simulations, whereas the line is the application of the approximate formula in equation \ref{eq:eq1DMLR}. (d-e) The integrated ionic charge as a function of the lateral position for $c_0=$ 1 mM and 100 mM. The charge is distributed over a wide area $O(L)$ extending beyond the pore mouth when electroneutrality is broken ($c_0$=1 mM), but is more localized when electroneutrality is maintained ($c_0$=100 mM) . (e) Quantification of end effects for two different concentrations. The electrostatic potential is plotted as a function of the $z$ coordinate, evaluated at the center axis of the channel. End effects are not significant for isolated channels.     } 
\end{figure*}

\subsection{Approximate formulas}
While we compute the full results of the PB equation in 3D, we compare the results to mathematical simplified formulas. As outlined in more detail in Appendices B and C, we can reduce our partial differential equation system of the 3D PB and Laplace equations into ordinary differential equations with appropriate boundary conditions. For small potentials, we then linearize the equations and get simple analytical formulas for the number of ionic charges within the membrane. In order to quantify the extent of electroneutrality breakdown, we take the ratio for the integrated amount of charge within the pore and the integrated amount of charge on the pore walls: $\mid Q_\mathrm{in}/Q_\mathrm{out} \mid$. In the limit of electroneutrality, we get  $\mid Q_\mathrm{in}/Q_\mathrm{out} \mid\rightarrow 1$, whereas in the limit of complete electroneutrality breakdown, we get $\mid Q_\mathrm{in}/Q_\mathrm{out} \mid\rightarrow 0$.
\begin{figure*}[!htbp] \label{fig:fig3}
\includegraphics[width=1\linewidth]{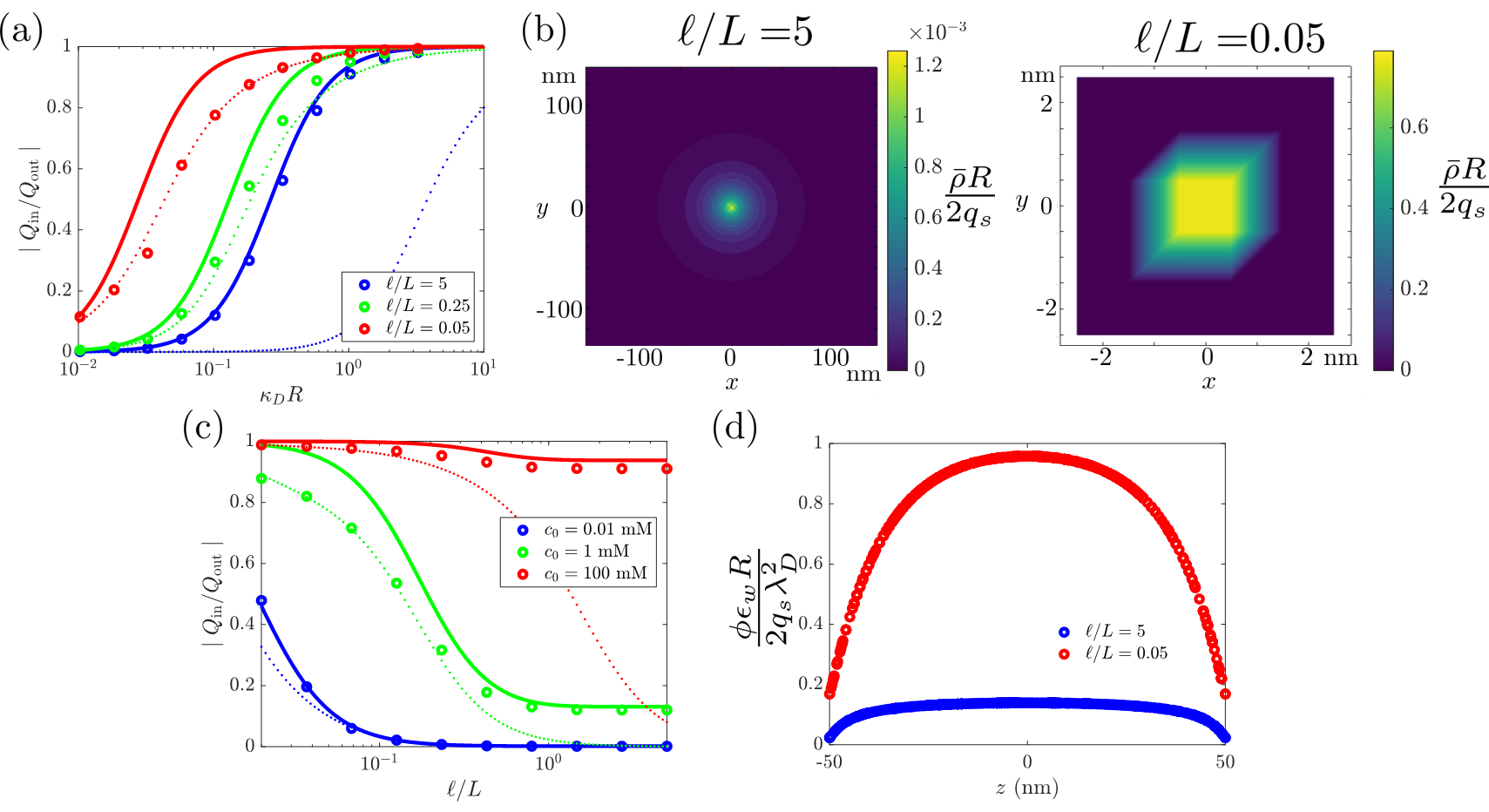}
\caption{Extent of electroneutrality breakdown for channels of different center-to-center separation distances on a lattice.
 (a) $\mid Q_\mathrm{in}/Q_\mathrm{out}\mid$ versus $\kappa_D R$ for $\ell/L=5$, $\ell/L=1$, $\ell/L=0.1$. (c) $\mid Q_\mathrm{in}/Q_\mathrm{out}\mid$ versus $\ell/L$ for $c_0=0.01, 1,100$ mM for $\ell/L=0.02$ to $\ell/L=0.5$. For both (a) and (c), the markers are the COMSOL simulations, whereas the solid lines are the application of the approximate formula in equation \ref{eq:eq1DMLR} and the dotted lines are the application of the approximate formula in equation \ref{eq:eqEndEffects}. (b) The integrated ionic charge as a function of the lateral position for $\ell/L=$ 5 and 0.05, with $c_0=$ 1 mM. The charge is distributed over a wide area when channels are isolated, but is localized when the channels are closely spaced and strongly interacting. (d) Quantification of end effects for two different lattice spacings with $c_0=$ 1 mM. The electrostatic potential is plotted as a function of the $z$ coordinate, evaluated at the center axis of the channel. End effects are significant when the channels are interacting.      } 
\end{figure*}

\textbf{No end effects:} Ignoring end effects, the inner potential can be solved for in terms of only the radial coordinate:
\begin{equation}
    \frac{\epsilon_w}{r}\frac{d}{dr}\left(r\frac{d \phi}{dr}\right)=2ec_0 \sinh\left(\frac{e\phi}{k_B T}\right)
\end{equation}
with boundary conditions given by:
\begin{equation} \label{eq:eqBCMLR}
    \frac{d\phi}{dr}(r=R)=\frac{q_s}{\epsilon_w}-\frac{\epsilon_m}{\epsilon_w}\frac{\phi(r=R)}{R M_{L/R}}, \quad \frac{d\phi}{dr}\left(r=\frac{\ell}{2}\right)=0
\end{equation}
where the constant $M_{L/R}$ is:
\begin{equation}\label{eq:eqMLR}
    M_{L/R}=\log \left(\frac{2L}{\pi R}\right)-\gamma+\frac{\mathrm{K}_1(\frac{\pi\ell}{2L})}{\mathrm{I}_1(\frac{\pi\ell}{2L})}.
\end{equation}
with $\gamma=0.577$ signifying the Euler-Mascheroni constant. Here, we assume circular shape of a unit cell for analytical simplicity, and the derivation is presented in Appendix B. Linearizing the equations and solving, we find the folllowing relationship for the amount of charge inside versus the amount of charge fixed on the pore walls:
\begin{equation}\label{eq:eq1DMLR}
    \left\rvert \frac{Q_\mathrm{in}}{Q_\mathrm{out}}\right\rvert=\frac{1}{1+\frac{\epsilon_m}{\epsilon_w}\frac{2}{{\kappa_D}^2 R^2 M_{L/R}}}
\end{equation}

\begin{figure*}[!htbp] \label{fig:fig4}
\includegraphics[width=1\linewidth]{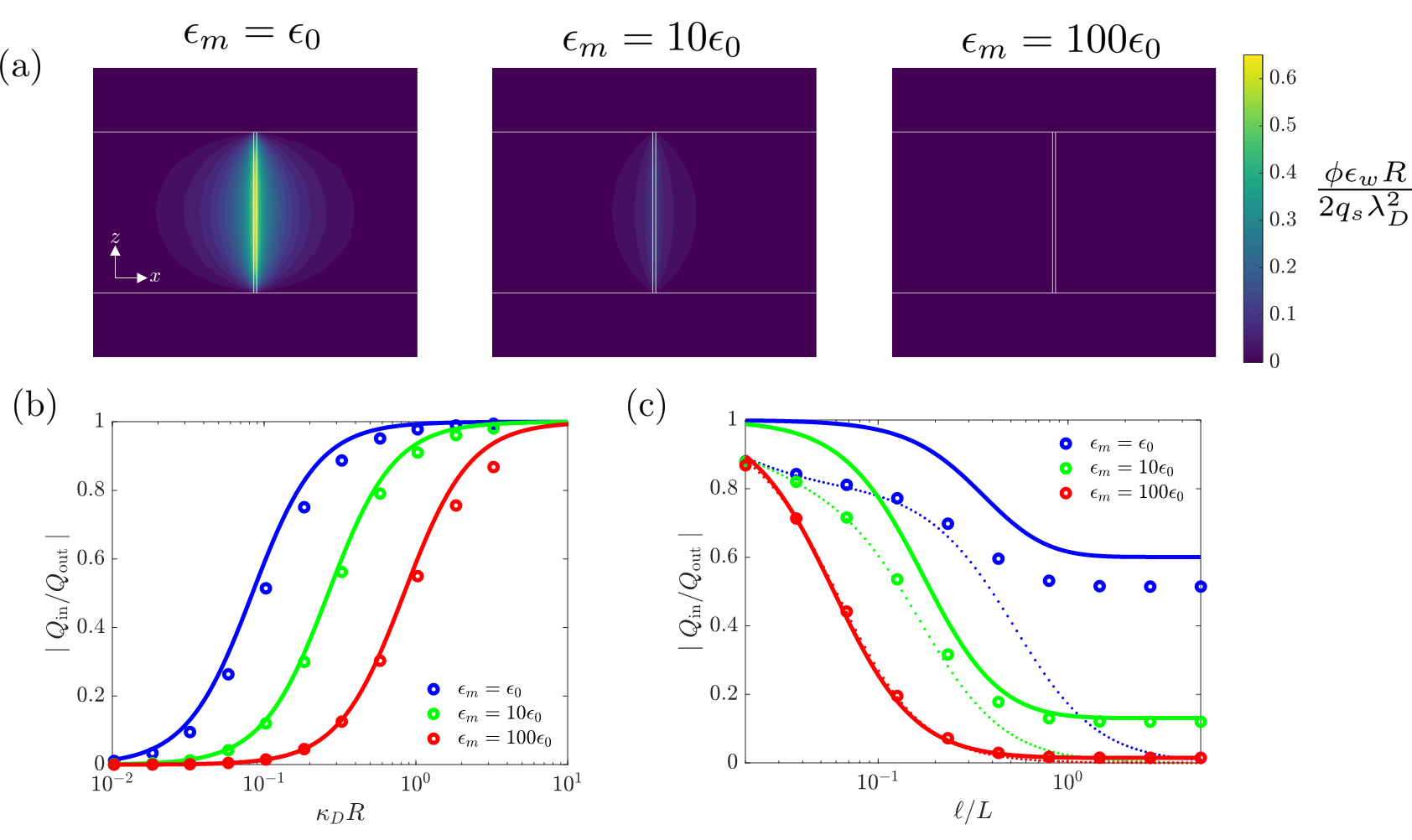}
\caption{Role of dielectric mismatch on extent of electroneutrality breakdown. (a)Results for isolated pore with same properties as in Fig. 1 (e), but with $\epsilon_m=\epsilon_0$, $\epsilon_m=10\epsilon_0$, and $\epsilon_m=100\epsilon_0$. (b) $\mid Q_\mathrm{in}/Q_\mathrm{out}\mid$ versus $\kappa_D R$ for varying $\epsilon_m$. (b) $\mid Q_\mathrm{in}/Q_\mathrm{out}\mid$ versus $\ell/L$ with $c_0=1$ mM for varying $\epsilon_m$.For both (b) and (c), the markers are the COMSOL simulations, whereas the solid lines are the application of the approximate formula in equation \ref{eq:eq1DMLR}  and the dotted lines are the application of the approximate formula in equation \ref{eq:eqEndEffects}.   }
\end{figure*}

Looking at the form of the above equation, one recognizes that electroneutrality breakdown is promoted as $\kappa_D R\rightarrow 0$. Furthermore, the effect has a weak dependence on the length of non-interacting nanopores, since the length appears logarithmically in $M_{L/R}$. However, the electrostatic interactions of the cylinders with each other can cause a return to electroneutrality as $\ell/L\rightarrow 0$. Furthermore, the amount of charge within the pore is strongly controlled by the dielectric constant of the membrane (and not necessarily the inner pore dielectric constant).

\textbf{No radial variations:} On the other hand, when channels become strongly interacting ($\ell/L\rightarrow 0$), the potential variations happen principally in the axial direction along the pore, governed by the dimensionless ratio $\lambda_D/L$. In order to capture the end effects of the channel, which become important on the scale of the Debye length, we write down a homogenized equation for the potential in linear response, which neglects radial variations in the potential by integrating over the lateral dimensions of a unit cell:
\begin{equation}\label{eq:eqAvgPois}
    \bar{\epsilon}\frac{d^2\phi}{dz^2}=-\frac{2 q_s}{R}+2ec_0\sinh\left(\frac{\phi e}{k_B T}\right)
\end{equation}
\begin{equation}\label{eq:eqAvgPoisPerm}
    \bar{\epsilon}=\epsilon_w-\epsilon_m+\epsilon_m\left(\frac{\ell^2}{\pi R^2}\right)
\end{equation}
The linear response boundary conditions are given by:
\begin{equation} \label{eq:eqAvgPoisBC}
    \frac{d\phi}{dz}\left(z=0,L\right)=\pm\frac{\epsilon_w}{\bar{\epsilon}}\frac{\ell^2}{\pi R^2}\frac{\phi\left(z=0,L\right)}{\lambda_D}.
\end{equation}
Solving the set of equations, we get the following fraction of charge inside the pore.
\begin{equation} \label{eq:eqEndEffects}
    \left\rvert \frac{Q_\mathrm{in}}{Q_\mathrm{out}}\right\rvert=\frac{1-(\gamma-p)\tanh(1/\gamma)}{1+p\tanh(1/\gamma)}
\end{equation}
with constants
\begin{equation}
    \gamma=2\sqrt{\frac{\bar{\epsilon}}{\epsilon_w}}\frac{\lambda_D}{L}, \quad p=\sqrt{\frac{\bar{\epsilon}}{\epsilon_w}}\frac{\pi R^2}{\ell^2}
\end{equation}

For strongly interacting channels, the electroneutrality condition is controlled by the ratio of $\lambda_D/L$. As $\lambda_D/L\rightarrow 0$, the system returns to overall electroneutrality within the pore.

Following the theoretical argument, a comprehensive set of numerical simulations is presented to validate the above formulas in their regime of validity, focusing on the linear regime with small but finite surface charge, $q_s\rightarrow 0$. The standard conditions chosen for the simulations, unless otherwise stated, are a pore length of 100 nm, a pore radius of 1 nm, a membrane permittivity of $\epsilon_m=10\epsilon_0$, an electrolyte permittivity of $\epsilon_w=80\epsilon_0$. Finally, a comparison is also made to the nonlinear solution of the full 3D equations. 

The results are presented with a number of dimensionless numbers, so a reader can easily interpret the plots. First, the dimensionless number $\kappa_D R$ indicates the extent of double layer overlap within the channel. $\kappa_D R\rightarrow0$ indicates strong double layer overlap while $\kappa_D R\rightarrow\infty$ indicates thin double layers relative to the pore radius. $\ell/L$ is the ratio of the center to center spacing between channels to the length of the channels. In plotting, the potential is normalized by the charge per unit length of a cylinder:
\begin{equation}\label{eq:eqDimPot}
    \tilde{\phi}=\frac{\phi \epsilon_w R}{2 q_s \lambda_D^2}
\end{equation}
One can roughly interpret these graphs as $\tilde\phi\approx 1$ means local electroneutrality in a give cross section of the pore, and $\tilde\phi\approx 0$ as local electroneutrality within the pore. Further, the depth-integrated charge density is plotted as a function of lateral position, to illustrate the extent of screening charges at the membrane/reservoir interfaces:
\begin{equation}
    \bar{\rho}_e(x,y)=\frac{\int \rho_e(x,y,z) dz}{L}.
\end{equation}

\section{Results and Discussion}
First, Fig. 1(e) summarizes the main trends seen in the simulations, where the potential is plotted as a function of position around the pore, as the spacing between pores is modified. When channels are isolated ($\ell=129$ nm), the potential within the channel is fairly constant. The variations in the potential in the radial direction are dominant. For the parameters chosen ($\kappa_D R=0.1$, $c_0=$ 1 mM), the amount of charge within the channel is only 12\% of the charge on the pore walls. However, as the size of a unit cell is reduced ($\ell=17$ nm) the variations in the potential become dominant in the axial direction, and edge effects become more pronounced. Furthermore, the potential does not vary significantly in the radial direction from the channel center axis.  The close channel spacing also limits the amount of field that can escape out of the pore, meaning that more charges are present within the channel, 43\%. 

\begin{figure*}[!htbp] \label{fig:fig5}
\includegraphics[width=1\linewidth]{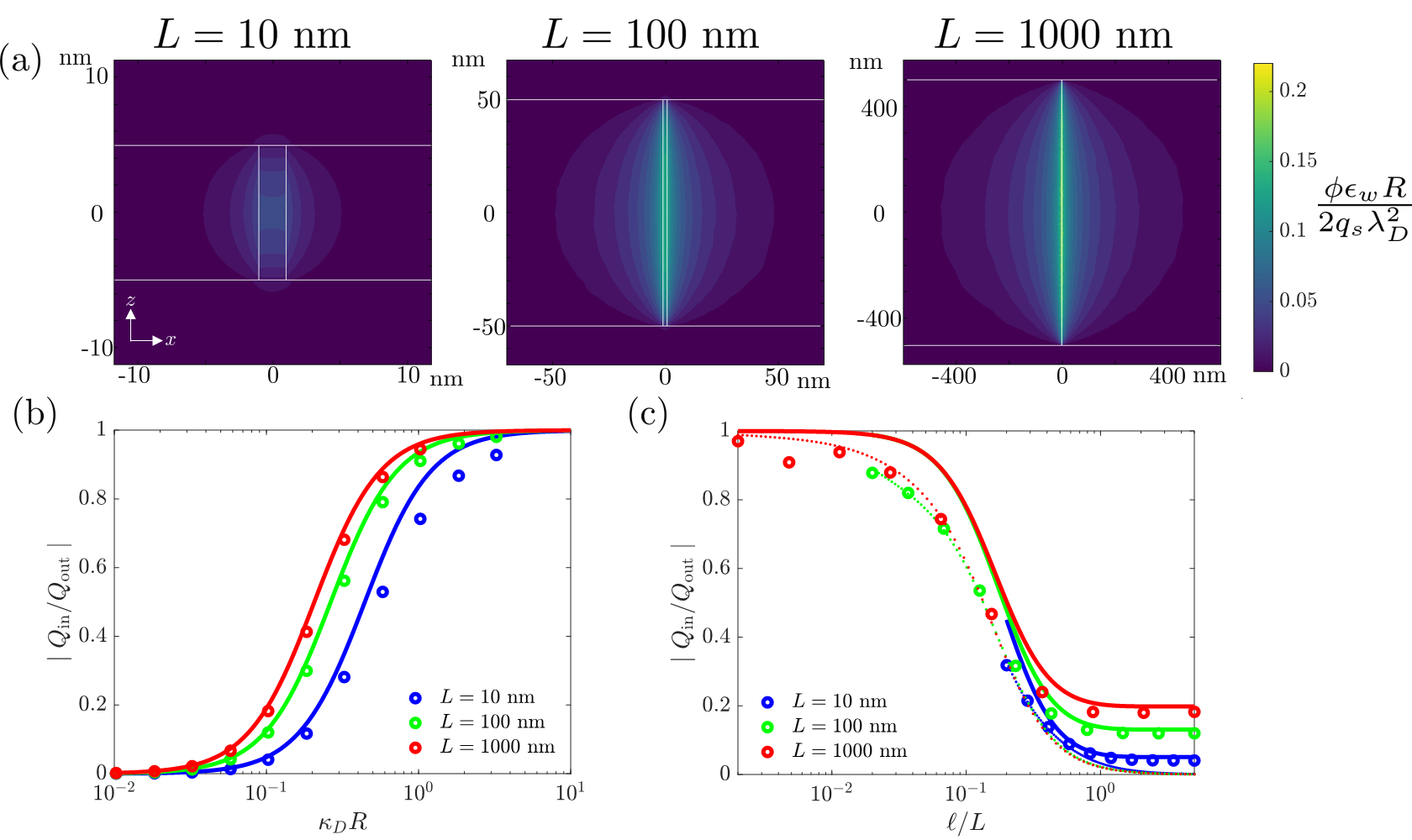}
\caption{Role of aspect ratio on extent of electroneutrality breakdown. (a)Results for isolated pore with same properties as in Fig. 1 (e), but with $L=10$ nm, $L=100$ nm, and $L=1000$ nm, also with $c_0=$ 1 mM. (b) $\mid Q_\mathrm{in}/Q_\mathrm{out}\mid$ versus $\kappa_D R$ for varying $L$, with $\ell/L=5$. (b) $\mid Q_\mathrm{in}/Q_\mathrm{out}\mid$ versus $\ell/L$ with $c_0=1$ mM for varying $L$. For both (b) and (c), the markers are the COMSOL simulations, whereas the solid lines are the application of the approximate formula in equation \ref{eq:eq1DMLR} and the dotted lines are the application of the approximate formula in equation \ref{eq:eqEndEffects}.   }
\end{figure*}

The trends in the potential profiles as a function of the salt concentration are shown in Fig. 2. Fig. 2(a) corresponds to $\kappa_D R=0.1$ or $c_0$=1 mM, and 2(b) corresponds to $\kappa_D R=1$ of $c_0=$ 100 mM, illustrating that electroneutrality is restored as the concentration is increased. The solid line in Fig. 2(c) given by equation \ref{eq:eq1DMLR} coincides quite closely with the results of the continuum simulations for the isolated channels. For these cases, neglecting end effects is a reasonable approximation for determining the number of charges within the pore, which is confirmed via the plot of the potential as a function of position along the center axis of the pore in Fig. 2(f). \color{black} The `plumes' of screening charge near the pore mouth are not immediately visible in the plots of the electrostatic potential, since the charges are far less concentrated outside of the pore. We can ascertain the extent of the screening at the membrane interfaces by analyzing the depth-integrated charge density. \color{black} In Fig. 2(d-e), the depth-integrated charge density, $\bar{\rho}_e$, is plotted as a function of the $x$ and $y$ coordinate. When electroneutrality is broken within the pore, the screening charge is distributed over the membrane surface over a distance that is on the order of the channel length, $L$. However, if electroneutrality is maintained, the screening charge is localized within the channel, and does not extend very far beyond the pore mouth. 

\begin{figure*}[!htbp] \label{fig:fig6}
\includegraphics[width=0.9\linewidth]{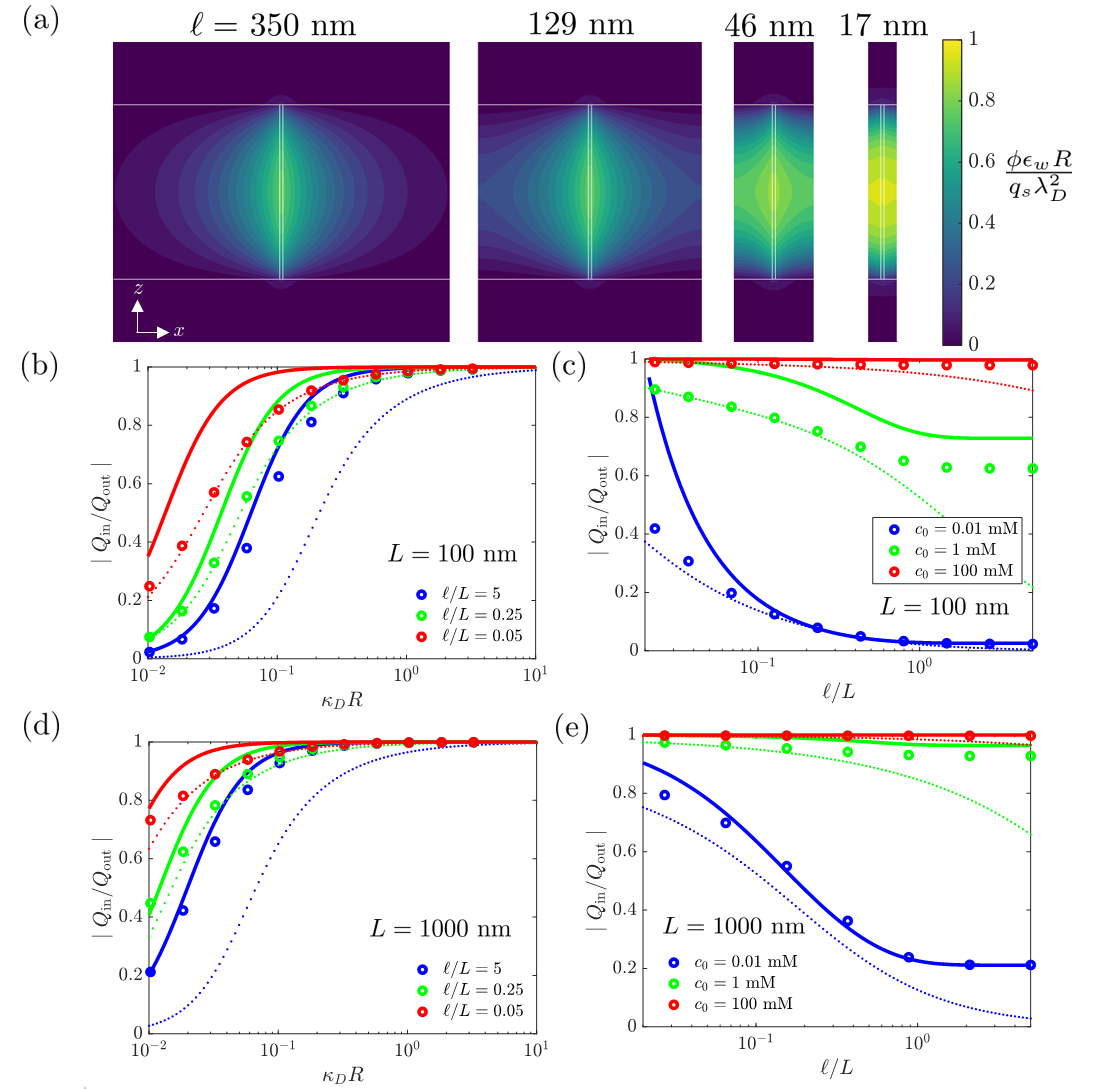}
\caption{Role of dimensionality of confinement by inspecting a slit pore geometry. (a)Results for isolated pore with same properties as in Fig. 1 (e), but for a slit pore. (b) $\mid Q_\mathrm{in}/Q_\mathrm{out}\mid$ versus $\kappa_D R$ for varying $\ell/L$. (c) $\mid Q_\mathrm{in}/Q_\mathrm{out}\mid$ versus $\ell/L$ for varying $c_0$. (d-e) The same as (b-c) but with $L=1000$ nm.  For (b), (c), (d), and (e), the markers are the COMSOL simulations, whereas the solid lines are the application of the approximate formula in equation \ref{eq:eq2DMLR} and the dotted lines are equation \ref{eq:eqEndEffects} with constants given by equations \ref{eq:eq2Dconst1} and \ref{eq:eq2Dconst2}.   } 
\end{figure*}

Next, we isolate the influence of channels interacting through the membrane domain, via modifying the ratio $\ell/L$ in Fig. 3. As channels are closer together, $\ell/L\rightarrow 0$, the amount of charge within the channel is increased closer towards electroneutrality. We also see a clear difference between the predictions neglecting end effects (equation \ref{eq:eq1DMLR}) in solid lines and the predictions neglecting radial variations (equation \ref{eq:eqEndEffects}) in dotted lines. As $\ell/L\ll1$, the predictions of the model neglecting radial variations become superior. However, when $\ell/L$ is 1 or greater, the predictions of the model that neglects end effects are superior. Such a result is expected, given the illustrative example in Fig. 1(e), where the importance of end effects are visible on the plot when channels are close together. For further confirmation, the magnitude of the electric field is plotted as a function of position for varying channel spacings in Fig. S2, exhibiting the importance of end effects for closely spaced channels. In Fig. 3(d), the potential plotted as a function of position on the center axis of the nanopore shows significant end effects for closely spaced channels ($\ell/L$=0.05). Also, the closely spaced channels localize the screening charge near the pore mouth, as evidenced by the depth-averaged charge density in Fig. 3(b).

\begin{figure*}[!htbp] \label{fig:fig7}
\includegraphics[width=1\linewidth]{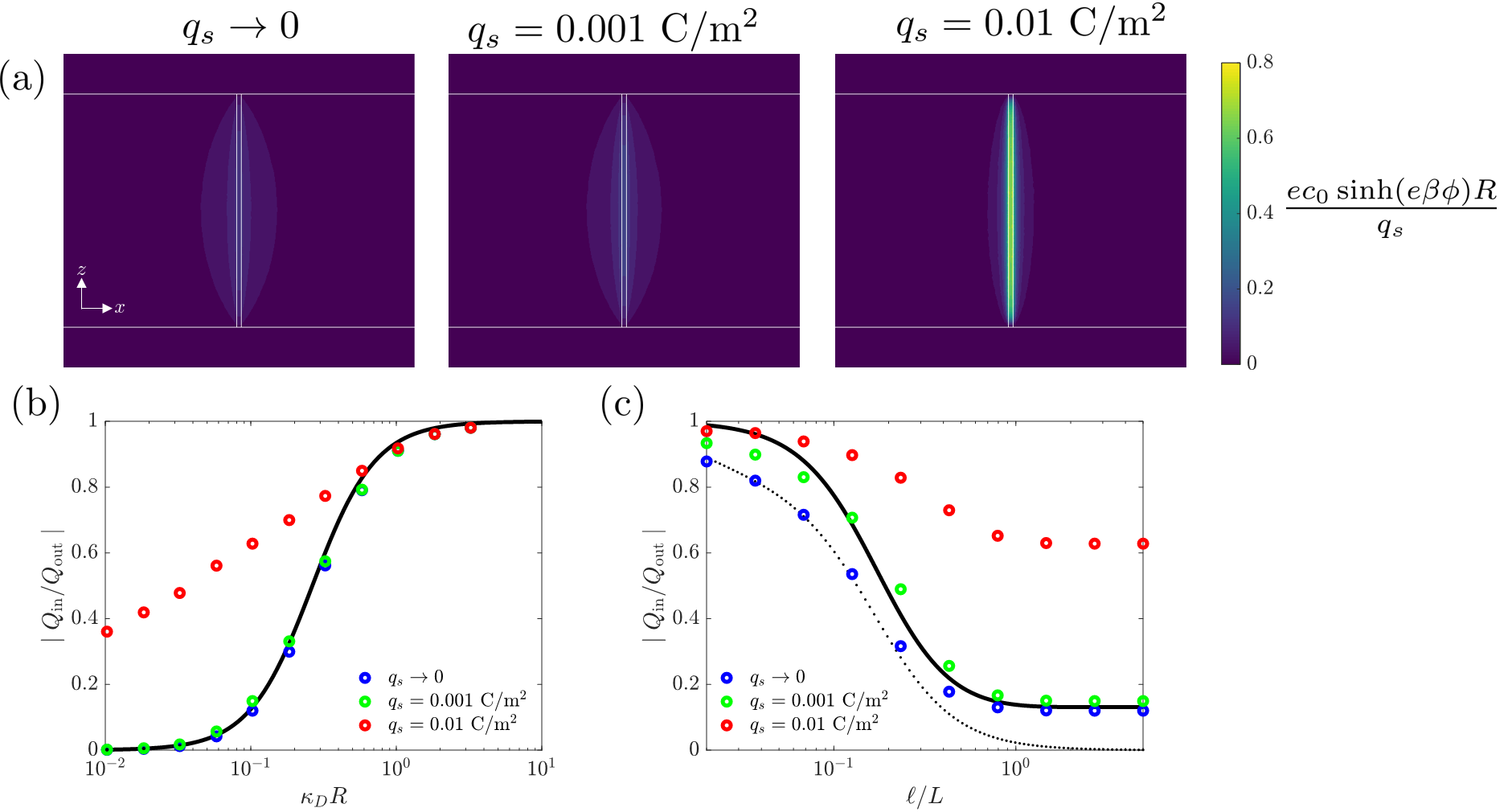}
\caption{Role of nonlinearity on extent of electroneutrality breakdown. (a)Results for isolated pore with same properties as in Fig. 1 (e), but with $q_s\rightarrow 0$, $q_s=0.001$ C/m$^2$, and $q_s=0.01$ C/m$^2$ with $c_0=$ 1 mM. (b) $\mid Q_\mathrm{in}/Q_\mathrm{out}\mid$ versus $\kappa_D R$ for varying $q_s$. (b) $\mid Q_\mathrm{in}/Q_\mathrm{out}\mid$ versus $\ell/L$ with $c_0=1$ mM for varying $q_s$. For both (b) and (c), the markers are the COMSOL simulations, whereas the solid line is the application of the approximate formula in equation \ref{eq:eq1DMLR} and the dotted line is the application of the approximate formula in equation \ref{eq:eqEndEffects}.   } 
\end{figure*}

The interactions of channels has a profound impact on the selectivity of a multipore dielectric membrane. The ensemble interactions can lead to a return to electroneutrality in strongly confined systems, leading to more charge selectivity and higher channel conductance. The extent of electrokinetic couplings are also maximized when the amount of ionic charge within the pore is higher ~\cite{peters2016analysis, VanderHeyden2006}. Such a design principle could be used to promote higher electrokinetic conversion efficiencies in ``blue energy'' harvesting of salinity gradients ~\cite{POST2007218, He2017}.

Another important design parameter is the dielectric constant of the membrane matrix. In Figure 4, we examine the influence of the membrane dielectric constant on the extent of electroneutrality breakdown. As the membrane dielectric constant decreases, the system moves closer towards electroneutrality in the pore. Again, the predictions from equations \ref{eq:eq1DMLR} and \ref{eq:eqEndEffects} seem to describe the data quite well within their respective realms of validity. 

A critical question remains to understand the influence of the aspect ratio on the extent of electroneutrality breakdown, the subject of Figure 5. Again observing equation \ref{eq:eq1DMLR}, for an isolated channel, the electroneutrality is enforced as $\log(L/R)\rightarrow\infty$. Therefore, the electroneutrality breakdown is only weakly affected by the length of the channel. As shown in Figure 5(b), we see only small shifts in the extent of electroneutrality breakdown with increasing channel length, $L$. Furthermore, the channels exhibit similar behavior with varying channel spacing, $\ell$. Such results arise from the exponentially long screening length in one dimensional confinement~\cite{levy2020breakdown}. The weak dependence of the extent of electroneutrality on the aspect ratio is a hallmark of electroneutrality breakdown in 1D channels.

As a point of comparison, it is instructive to perform the same analysis for two dimensional confinement, or slit pore geometry. Here, we perform analogous continuum simulations to the 1D confinement case. In Figure 6, we have simulated a slit pore with width $2R$ and length $L$, with channel center to channel center distances of $\ell$. We compare the simulation results to two analytical formulas, one where we have neglected end effects, and one where we have neglected normal variations in the potential to the pore walls (again termed ``neglecting `radial' variations''), similar to the analysis for 1D confinement, as outlined below.


The extent of electroneutrality breakdown in 2D confinement, \textbf{ignoring edge effects}, is:
\begin{equation} \label{eq:eq2DMLR}
    \left\rvert \frac{Q_\mathrm{in}}{Q_\mathrm{out}}\right\rvert=\frac{1}{1+\frac{\epsilon_m}{\epsilon_w}\frac{1}{{\kappa_D}^2 R^2 M_{L/R}}}
\end{equation}
with modified constant $M_{L/R}$ given by:
\begin{equation}
    M_{L/R}=\frac{L}{\pi R}\coth\left(\frac{\pi}{L}(\ell/2-R)\right)
\end{equation}
Observing the form of the 2D $M_{L/R}$ in this scenario, it is apparent that electroneutrality is enforced as $L/R\rightarrow\infty$. This means that electroneutrality will be much more strongly upheld in 2D confinement compared to 1D confinement.

When $\ell/L$ becomes smaller, the potential variations in the axial direction of the 2D slit pore become dominant, similar to 1D confinement. The extent of electroneutrality breakdown \textbf{neglecting `radial' variations} is given identically by equation \ref{eq:eqEndEffects}, but with constants
\begin{equation}\label{eq:eq2Dconst1}
    \gamma=2\sqrt{\frac{\bar{\epsilon}}{\epsilon_w}}\frac{\lambda_D}{L}, \quad p=\left(\frac{\bar{\epsilon}}{\epsilon_w}\frac{4 R^2}{\ell^2}\right)^{1/2}
\end{equation}
where 
\begin{equation}\label{eq:eq2Dconst2}
    \bar{\epsilon}=\epsilon_w-\epsilon_m+\epsilon_m\left(\frac{\ell}{2 R}\right).
\end{equation}
The progression towards electroneutrality breakdown for 2D confinement is similar to 1D confinement when the channel spacing is close together. 

In Fig. 6(b-e), we see that the propensity towards electroneutrality is much stronger in 2D confinement than in 1D confinement. We find that, similar to 1D confinement, electroneutrality is promoted when channels are strongly interacting. We also find that the extent of electroneutrality breakdown is extremely sensitive the to the length of the channel domain, especially as compared to 1D confinement, as emphasized by comparing Fig. 6(d-e) to Fig. 6(b-c). Note that the normalization of the potential in Fig. 6(a) is adjusted from the definition in equation \ref{eq:eqDimPot} due to the difference in geometry.

So far, we have examined the linear regime with small but finite values of $q_s$. Evidently, the nonlinearity in the equations will affect the validity of our approximations in equations \ref{eq:eq1DMLR} and \ref{eq:eqEndEffects}. In Fig. 7, we show that as the charge density is increased into the nonlinear regime, the system moves closer towards electroneutrality. The approximations we derived earlier are insufficient to describe the extent of electroneutrality breakdown in the nonlinear limit.  It is particularly difficult to derive analytical approximations in the nonlinear limit, so we do not explore such approximations here. Even so, for isolated channels, the electroneutrality breakdown emerges as a function of the ratio of the Gouy-Chapman length to the pore radius in this limit~\cite{levy2020breakdown}.

Finally, we preview the possible manifestations of electroneutrality breakdown that could be observed in experiments: single channel conductance and transference number, using the simplified formulas in equation \ref{eq:eq1DMLR} and equation \ref{eq:eqEndEffects}. We assume a KCl solution with fixed and equal mobilites, equal to the bulk value of $D=D_+=D_-=2\times10^{-9}$ m$^2$/s. For an uncharged pore, the fraction of current carried by each ion would be 50\%. However, in this case, we assume that the pore walls are negatively charged with a value of $q_s=-0.001$ C/m$^2$. Therefore, we approximate the anion and cation concentrations within the channel as:
\begin{equation}
    c_+=c_0+\frac{2 \mid q_s\mid}{e R}\frac{\mid Q_\mathrm{in}\mid}{\mid Q_\mathrm{out}\mid}, \quad c_-=c_0
\end{equation}
We approximate the overall channel conductance as:
\begin{equation}
    G= \frac{2 \pi D R^2  e^2 c_0}{k_B T L}+\frac{2  \pi D R e \mid q_s\mid}{ k_B T L}\frac{\mid Q_\mathrm{in}\mid}{\mid Q_\mathrm{out}\mid}
\end{equation}
which can be rendered dimensionless:
\begin{equation}
    \tilde{G}= G/\left(\frac{2 \pi D R^2  e^2 c_\mathrm{ref}}{k_B T L}\right)=\frac{c_0}{c_\mathrm{ref}}+\frac{ \mid q_s\mid}{ R e c_\mathrm{ref}}\frac{\mid Q_\mathrm{in}\mid}{\mid Q_\mathrm{out}\mid}
\end{equation}
where $c_\mathrm{ref}$ is arbitrarily chosen to be 1 mM. 
The corresponding cation transference number, or fraction of current carried by the cation, is:
\begin{equation}
    t_+= \left(c_0+\frac{2 \mid q_s\mid}{e R}\frac{\mid Q_\mathrm{in}\mid}{\mid Q_\mathrm{out}\mid}\right)/\left(2 c_0+\frac{2 \mid q_s\mid}{e R}\frac{\mid Q_\mathrm{in}\mid}{\mid Q_\mathrm{out}\mid}\right).
\end{equation}
The dimensionless conductance and cation transference number are shown in Fig. 8, for a channel with the sample parameters: $L=$ 100 nm, $R=$ 1 nm, $\epsilon_m=10\epsilon_0$,  $\epsilon_w=80\epsilon_0$, and $T=$ 300 K. The plateau in conductance does not occur when electroneutrality breakdown is present. The decrease in conductance at low concentration has been experimentally observed in Refs. ~\cite{secchi2016scaling} and ~\cite{yao2019strong}, but the effects were ascribed to surface reactions and electrokinetic coupling, respectively. Here, the large resistance through the pore is expected at low concentration due to electroneutrality breakdown, since fewer counterions are present as charge carriers. As channels become closely spaced or strongly interacting, their behavior returns to the plateau behavior. The presence of electroneutrality breakdown does not rule out the previous explanations for deviations from the conductance plateau at low concentrations. However, the charge regulation reactions and electrokinetic effects might be less sensitive to the interactions of the pores. In addition to electroneutrality breakdown, experiments might also include resistances incurred from microchannel domains that connect to the nanopore.\cite{green2016interplay, green2014effect} In terms of the transference number, electroneutrality breakdown at low concentrations leads to a cation transference number that does not saturate at $t_+=1$, again due to the reduction in screening counterions. Another practically significant quantity is the capacitance of a conducting nanotube embedded in a membrane dielectric medium, which is explored in the Supplemental Information. Electroneutrality breakdown can be used to increase the effective capacitance per unit pore area at low ionic concentrations. However, the capacitance per total membrane area and per total membrane volume do not benefit from electroneutrality breakdown, since the dense channel spacing reduces the effectiveness of electroneutrality breakdown.

\begin{figure}[b] \label{fig:fig8}
\includegraphics[width=0.9\linewidth]{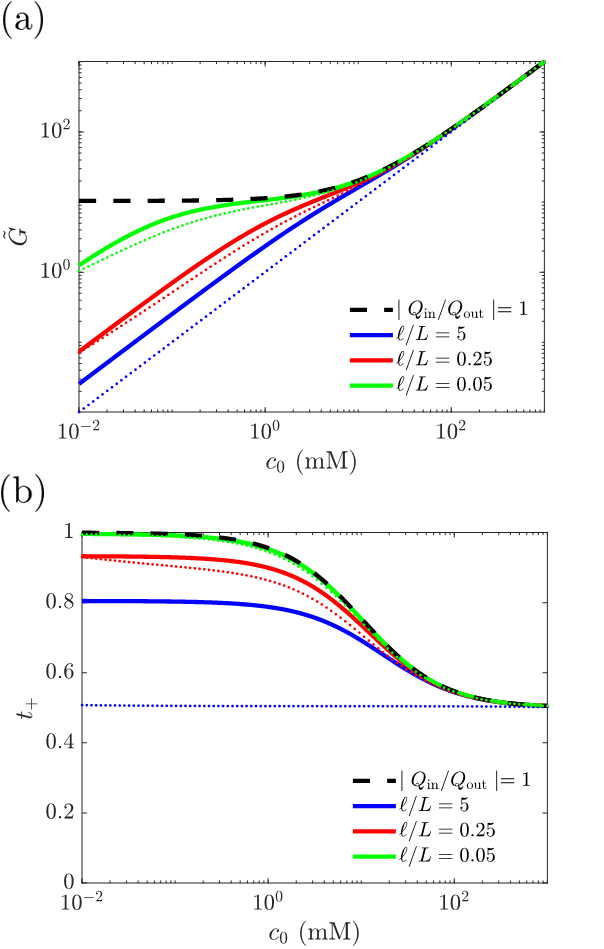}
\caption{ The dimensionless conductance (a) and cation transference number  (b) through a negatively charged nanochannel for varying channel separation distances. The solid lines are the predictions using equation \ref{eq:eq1DMLR} and the dotted lines are the predictions using equation \ref{eq:eqEndEffects}. The dotted lines are good predictors of the extent of electroneutrality breakdown at small $\ell/L$, but fail at large $\ell/L$, e.g. the blue dotted lines.  The plateau in conductance at low concentration is only apparent when $\mid Q_\mathrm{in}/Q_\mathrm{out}\mid=1$. As the ratio $\ell/L$ decreases, the system moves closer to electroneutrality.  When electroneutrality is broken, the cation transference number does not saturate to 1 at low concentration.}
\end{figure}

Nanoconfined domains are most ubiquitous in biological membranes, where protein channels selectively conduct specific ions. The selectivity filter in ion channels are at the molecular scale, with a radii on the order of single angstroms ~\cite{malasics2009protein}. In our model thus far, the radius of the channel signifies the accessible area for the \textit{ion centers}, which becomes negligible in true molecular confinement, $R\rightarrow 0$.   Therefore, we expect electroneutrality breakdown to be critical in describing the conduction of ions through protein channels. In the absence of an electroneutrality constraint, molecular separation in the selectivity filter would instead rely on specific chemical interactions or energy barriers to enter the pore ~\cite{EPSZTEIN2019316}. Closely spaced channels on the order of the membrane thickness $\sim 10$ nm could interact electrostatically with each other, leading to ensemble gating and ion conduction events.

The electrokinetic coupling and electrokinetic conversion from salinity gradients or pressure gradients are most effective and efficient in the regime of strong double layer overlap. Electroneutrality breakdown can adversely impact the expected performance of such a process. A large density of channels is not only desired for higher flux membranes, but to eliminate the possibility of electroneutrality breakdown at low concentration. 

In unstructured charged nanoporous media, such as porous rock or polymer membranes, a large network of charged pores are connected and interact strongly with each other. Applications include desalination, ionic separations, and oil recovery~\cite{Dydek2013, deng2015water, schlumpberger2015scalable, alkhadra2019continuous, conforti2020continuous, alkhadra2020small}. In an interconnected porous medium consisting of closely spaced pores, a similar homogenized model to equation \ref{eq:eqEndEffects} should be pursued ~\cite{schmuck2015homogenization}. If the medium length scale is large relative to the Debye length, then electroneutrality will be maintained.

With our emphasis on the physics of electroneutrality breakdown, we have neglected other chemical mechanisms, such as charge regulation~\cite{trefalt2016charge, behrens2001charge, lin2019modulation, lin2019electrodiffusioosmosis}. Charge regulation, or reactions to form or neutralize surface charge, is certainly occurring at the pore walls-- after all, the origin of surface charge is a result of charge adsorption to the interface. Even if charge regulation is present, the effect of electroneutrality breakdown should still be important. In order to delineate from a chemical mechanism, one convincing evidence of electric field escape from a channel would be to observe differences in measureable quantities as a function of the spacing between channels. Only measurements that determine single channel conductance, transference number, or capacitance, as a function of the density of channels in a membrane could distinguish the electroneutrality breakdown mechanism from other competing chemical mechanisms. Furthermore, in Single Digit Nanopores channels, we can expect energy barriers associated with ions' dehydration to enter the channel. Energy barriers in the pore domain can be easily added to the model explored here, by changing the effective chemical potential for ions within the pore~\cite{levy2020breakdown}. Non-ideal effects ~\cite{bazant2009towards} such as packing effects ~\cite{kilic2007steric, Kornyshev2007} or electrostatic correlations~\cite{Bazant2011, storey2012effects,de2020continuum},are also not considered in this work.
\section{Conclusions}
The electroneutrality breakdown phenomenon is studied with an extensive set of continuum simulations. The results are shown to agree with simplifying analytical formulas within their regime of validity. Furthermore, the practical influence of electroneutrality breakdown on channel conductance and selectivity is discussed.

The experimental validation of screening in lower dimensions presents multiple competing mechanisms which can obscure the presence of electroneutrality breakdown. Even so, the set of simulation results presented here can guide researchers to isolate electroneutrality breakdown for multipore systems. One can expect wide variations of properties as a function of channel number density per area when electric fields enter into the membrane domain. As channels are placed closer together, they interact more strongly, changing the transport properties.

\section{Acknowledgements}
This research was supported as part of the Center
for Enhanced Nanofluidic Transport (CENT), an Energy
Frontier Research Center funded by the U.S. Department
of Energy, Office of Science, Basic Energy Sciences under
Award \# DE-SC0019112 .
JPD acknowledges support from the National Science
Foundation Graduate Research Fellowship under Grant
No. 1122374.

\bibliography{library}
\newpage
\appendix
\renewcommand\thefigure{S\arabic{figure}}
\setcounter{figure}{0}

\begin{figure*}[!htbp] 
\includegraphics[width=1\linewidth]{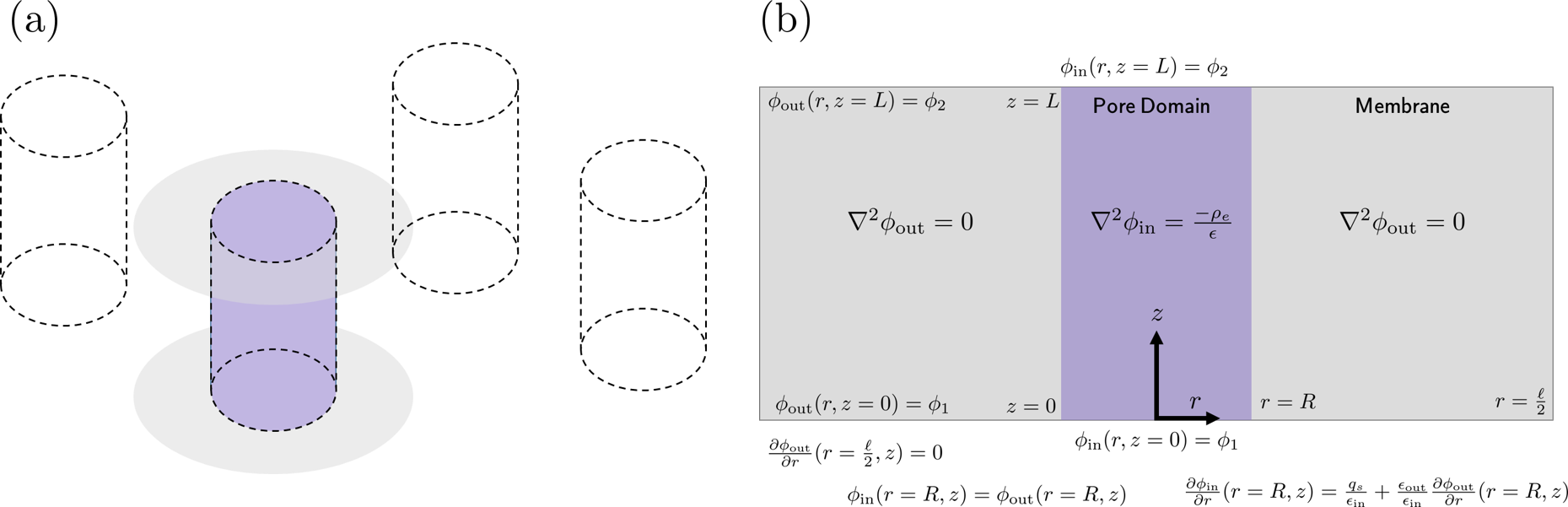}
\caption{The configuration of multiple channels in a membrane (a) is reduced to a single pore problem (b). The partial differential equation and boundary conditions are listed for the given single pore problem.  } 
\label{fig:figS1}
\end{figure*}
\section{Reduction of the PNP system into the PB system}
The PNP system of equations in the electrolyte domain that applies for transport measurements is given by coupling the Poisson equation to the Nernst-Planck equation for dilute species:
\begin{equation}
    \begin{split}
        &\epsilon_w\nabla^2\phi=-\rho_e\\
        &\mathbf{j_i}=-\frac{D_i c_i}{k_BT}\nabla\mu_i^{e}\\
        &\mu_i^{e}=k_B T \ln(c_i)+z_i e \phi
    \end{split}
\end{equation}
with steady state species conservation equations given by:
\begin{equation}
    \nabla\cdot \mathbf{j_i}=0.
\end{equation}
Since the channels we are considering are long and thin, we can neglect the change in flux in the normal direction to the pore walls ~\cite{peters2016analysis}, which is zero due to no penetration at the pore walls:
\begin{equation}
    j_{i,n}=-D_i \nabla_n c_i-\frac{D_i z_i e c_i}{k_BT}\nabla_n\phi=0
\end{equation}
Integrating with respect to the normal coordinate, we can define a Boltzmann distribution with:
\begin{equation}
    c_i(r,z)=c_{i,v}(z)\exp\left(\frac{-z_i e \phi}{k_B T}\right).
\end{equation}
Since both the open ends of the channel are fixed at approximately the reservoir concentration, we can choose the pre-exponential factor as $c_i,0$. At this point, we have reduced the dynamic Nernst-Planck equation into the equilibrium Poisson-Boltzmann equation. Driving forces applied to the system will linearly act upon the equilibrium configuration, as calculated in this paper. For driving forces that are not too large, we can expect the equilibrium conformation to also remain unperturbed by driving forces. Here, we have neglected the coupling to the Stokes equation, but it can be easily added to the analysis ~\cite{peters2016analysis}.

\begin{figure*}[!htbp] 
\includegraphics[width=1\linewidth]{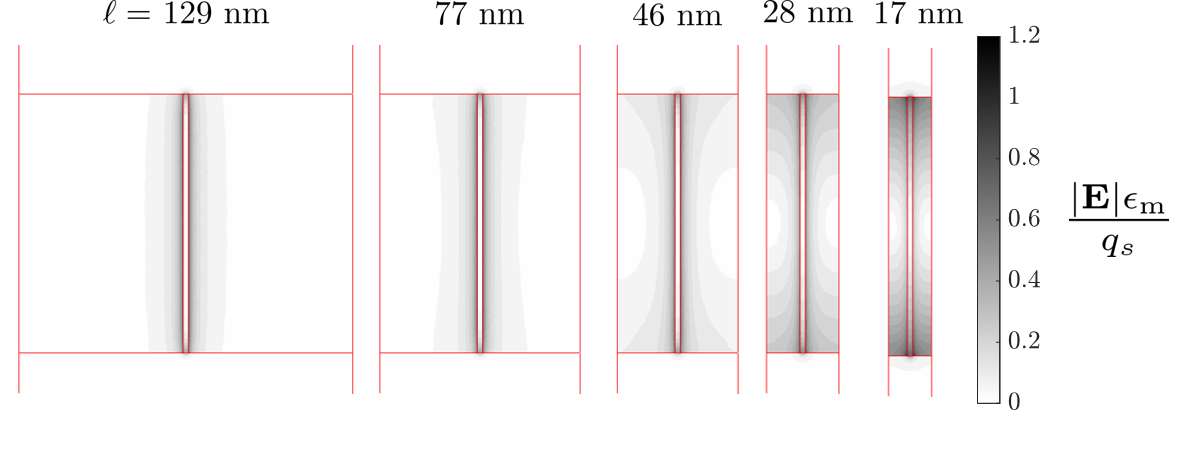}
\caption{ Electric field magnitude normalized by the surface charge as a function of position within a cross section of the membrane/channel system, for the same conditions as Fig. 1(e). As the unit cell becomes smaller and channels are closer together, the electric field lines are strongest at the entrance regions of the channel. End effects can be important in this regime. } 
\label{fig:figS3}
\end{figure*}

\section{Neglecting end effects}
First, we derive the approximate formulas for the number of ions within the channel by neglecting end effects, based on the work ~\cite{levy2020breakdown} but extended to a nanopore array configuration. In order to simplify the formulas, we consider a cylindrical unit cell that encapsulates a single nanochannel, as shown in Fig. \ref{fig:figS1}. 
In the membrane domain, because the Laplace equation is long range, we must account for the variations in the axial direction:
\begin{equation}
    \frac{1}{r}\frac{d}{dr}\left(r\frac{d\phi_\mathrm{out}}{dr}\right)+\frac{d^2\phi_\mathrm{out}}{d z^2}=0
\end{equation}
With homogeneous boundary conditions $\phi_1=\phi_2=0$, the solutions have the form:
\begin{equation}
\begin{split}
    &\phi_\mathrm{out}(z,r)=\sum_{n=1}^\infty A_n\left[\mathrm{K}_0(\lambda_n r)+\alpha_n I_0(\lambda_n r)\right]\sin(\lambda_n z) \\ &\alpha_n=\frac{\mathrm{K}_1(\frac{\lambda_n\ell}{2})}{\mathrm{I}_1(\frac{\lambda_n\ell}{2})}, \quad \quad \lambda_n=\frac{n\pi}{L}
\end{split}
\end{equation}
The normal derivative to the channel wall is:
\begin{equation}
    \frac{\partial\phi_\mathrm{out}}{\partial r}(z,r)=\sum_{n=1}^\infty A_n\left[-\lambda_n\mathrm{K}_1(\lambda_n r)+\alpha_n \lambda_n I_1(\lambda_n r)\right]\sin(\lambda_n z).
\end{equation}
Far from the electrolyte/membrane interface, the first eigenvalue dominates the solution. We can approximate each summation asymptotically by truncating to the first term:
\begin{equation}
    \begin{split}
         &\phi_\mathrm{out}(z,R)\sim A_1\left[\mathrm{K}_0(\lambda_1 R)+\alpha_1 I_0(\lambda_1 r)\right]\sin(\lambda_1 z) \\
         &\frac{\partial\phi_\mathrm{out}}{\partial r}(z,R)\sim A_1\left[-\lambda_1\mathrm{K}_1(\lambda_1 R)+\alpha_1 \lambda_1 I_1(\lambda_1 R)\right]\sin(\lambda_1 z)
    \end{split}
\end{equation}
Taking the limit as $R/L\rightarrow 0$, we get:
\begin{equation}
    \begin{split}
        &\phi_\mathrm{out}(z,R)\sim A_1\left[-\gamma-\log(\frac{\pi R}{2 L})+\alpha_1 \right]\sin(\lambda_1 z)\\
        &\frac{\partial\phi_\mathrm{out}}{\partial r}(z,R)\sim -\frac{A_1}{R}\sin(\lambda_1 z)
    \end{split}
\end{equation}
where $\gamma$ is the Euler constant. Taking the ratio of the potential and its derivative at the external radius, we get:
\begin{equation}
    \frac{\partial\phi_\mathrm{out}}{\partial r}(z,R)= \frac{-\phi_\mathrm{out}(z,R)}{R M_{L/R}}
\end{equation}
with $M_{L/R}$ defined by equation \ref{eq:eqMLR}. If we apply the above relationship into the electric flux boundary conditions, we get the boundary condition in equation \ref{eq:eqBCMLR}. Next, we reduce the Poisson-Boltzmann equation from a partial differential equation to an ordinary differential equation by neglecting the variation in potential in the electrolyte domain in the $z$ direction:
\begin{equation}
     \frac{\epsilon_w}{r}\frac{d}{dr}\left(r\frac{d \phi}{dr}\right)=2ec_0 \sinh\left(\frac{e\phi}{k_B T}\right).
\end{equation}
Solving the approximate boundary conditions in the linear response, we can arrive at the equation for the amount of charge in the pore relative to the charge on the pore walls, given in equation \ref{eq:eq1DMLR}.

\section{Neglecting radial variations}
As shown in Fig. \ref{fig:figS3}, when channels become closely spaced, the electric field is concentrated at the ends of the channel, meaning that end effects become more important. In order to turn our partial differential equation system into an ordinary differential equation system valid in this regime, we average over the $x$ and $y$ dimensions of the unit cell. Integrating the Poisson and Laplace equation system over $x$ and $y$, dividing by $\pi R^2$, and assuming no variations in $\phi$ in the $x$ or $y$ directions, we get equations \ref{eq:eqAvgPois} and \ref{eq:eqAvgPoisPerm}. Next, at linear response, we can relate the derivative of the potential at the interface to the value of the potential itself at the interface, since the potential should exponentially decay into the reservoir domain. Mathematically, the relationship is given in equation \ref{eq:eqAvgPoisBC}. Note that the approximations we have made here are only valid if the potential is constant in each $z$-slice of our simulation. Such an approximation is expected to fail at large channel separation distances, as shown throughout our results.

\section{Unit cell shape}
\begin{figure}[!htbp] \label{fig:figS2}
\includegraphics[width=1\linewidth]{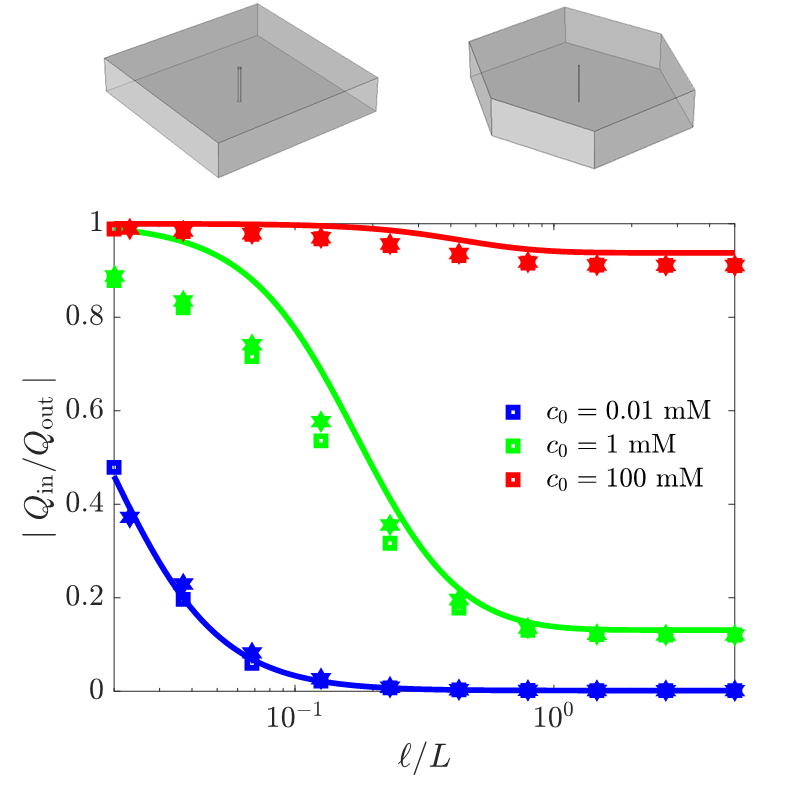}
\caption{ The extent of electroneutrality breakdown for two different unit cell shapes: square and hexagonal. The center to center distance between channels, $\ell$ is maintained the same between the channels. The conditions are identical to Fig. 3(c). As $\ell$ is varied, the hexagonal and square unit cells appear to have similar extents of electroneutrality breakdown. } 
\end{figure}

Here, we explore whether the shape of the unit cell affects the extent of electroneutrality breakdown. We find that the results for a hexagonal unit cell with the same center to center distance as a square unit cell exhibits the same extent of electroneutrality breakdown, as shown in Fig. \ref{fig:figS2}. While certain differences may arise by fluctuations from the lattice position or irregular lattice spacing of channels, we have provided some additional evidence that our square array is a good approximation for a variety of channel arrangements. 
\begin{figure*}[!htbp] 
\includegraphics[width=1\linewidth]{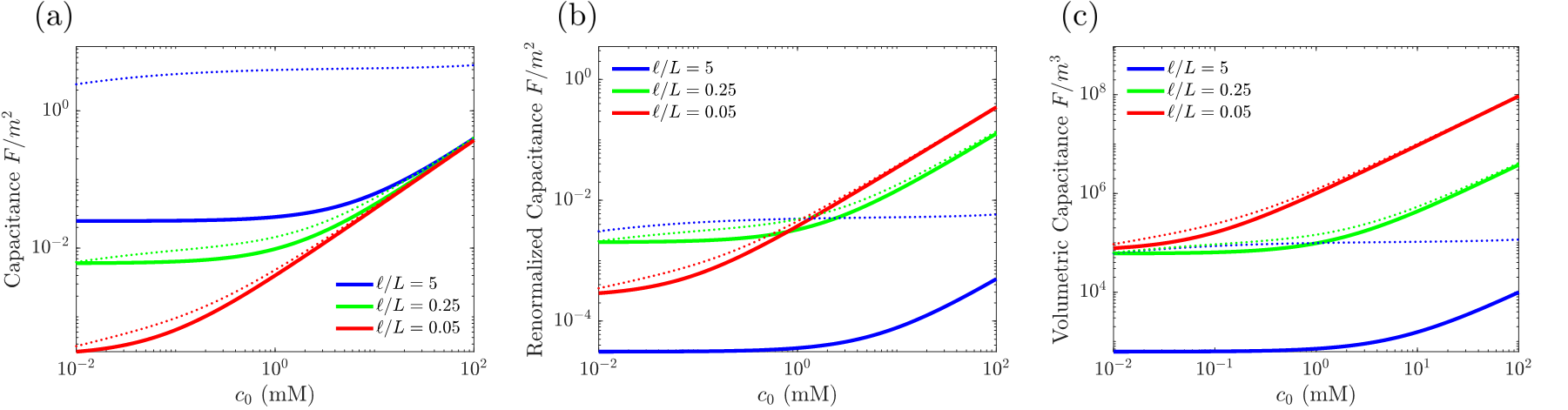}
\caption{The capacitance (a) normalized by the pore surface area  (b) normalized by the membrane/electrolyte surface area per unit cell and (d) normalized by the membrane/pore volume. In each plot, the separation distance between channels is varied, changing the extent of electroneutrality breakdown. The solid lines are the predictions using equation \ref{eq:eq1DMLR} and the dotted lines are the predictions using equation \ref{eq:eqEndEffects}. The dotted lines are good predictors of the extent of electroneutrality breakdown at small $\ell/L$, but fail at large $\ell/L$, e.g. the blue dotted lines.  } 
\label{fig:figS4}
\end{figure*}

\section{Capacitance}
As a final practical application of electroneutrality breakdown, we examine the capacitance of a nanopore that experiences electroneutrality breakdown. As a device, we imagine an electrically-connected nanopore embedded in a dielectric membrane. Scaled up, a set of nanopores is arranged in an array with center to center spacing of $\ell$, separated by a membrane dielectric domain. Using the same parameters as in Fig. 8, we calculate the differential capacitance within the channel at linear response:
\begin{equation}
    C=\frac{d q_s}{d\phi}=\frac{\epsilon_w R}{2\lambda_D^2}/\mid Q_\mathrm{in}/Q_\mathrm{out}\mid
\end{equation}
valid for strongly overlapping double layers $\kappa_D R\rightarrow 0$, and normalized by the surface area of the charged nanopore, $2\pi RL$. In Fig. \ref{fig:figS4}(a), the capcacitance is plotted, showing that for channels that have electroneutrality breakdown $\ell/L>1$ at low concentrations, that the capacitance saturates at low concentration. The value of the capacitance at low concentration is \textit{higher} for channels which exhibit electroneutrality breakdown. If we renormalize the capacitance by the total surface area of the membrane/electrolyte domain per unit cell, we arrive at the results in Fig. \ref{fig:figS4}(b). If we renormalize the capacitance by the volume of the membrane/pore system, we get the results in Fig. \ref{fig:figS4}(c). These renormalized capacitance values exhibit the opposite trend--namely, as channels are closer together ($\ell/L\rightarrow 0$), the cumulative surface area and volume decrease, leading to greater renormalized capacitance. Even though the single channels have more capacitance per pore surface area due to electroneutrality breakdown, the additional surface area or volume for the channels to be spaced far apart from each other leads to smaller renormalized capacitance. One could potentially design a membrane with significant density which still exhibits some electroneutrality breakdown, leading to better performance than a denser array of channels, as shown in the comparison between the green and red lines in Fig. \ref{fig:figS4}.

\end{document}